\documentclass[12pt]{article}
\newcommand{\ignore}[1]{}
\usepackage[english]{babel}
\usepackage{graphicx,rotating,epsfig}
\usepackage{footmisc}
\usepackage{amsmath}
\usepackage{amssymb}
\usepackage{dcolumn}   
\usepackage{graphicx}  
\usepackage{dcolumn}   
\usepackage{bm}        
\usepackage{amssymb}   
\newcommand{\MET}{\ensuremath{{\slash\kern-.7emE}_{T}}}
\newcommand{\pte}{p_T^{(e)}}
\newcommand{\mte}{M_T^{(e)}}
\newcommand{\ptmu}{p_T^{(\mu)}}
\newcommand{\mtmu}{M_T^{(\mu)}}
\newcommand{\METmu}{\MET^{(\mu)}}
\newcommand{\METe}{\MET^{(e)}}

\newcommand{\TeV}{\ensuremath{\mathrm{Te\kern -0.1em V}}}
\newcommand{\GeV}{\ensuremath{\mathrm{Ge\kern -0.1em V}}}
\newcommand{\MeV}{\ensuremath{\mathrm{Me\kern -0.1em V}}}
\def\GeVc2{\ensuremath{\mathrm{ Ge\kern -0.1em V }\kern -0.2em /c^2 }}

\newcommand{\MW}{\ensuremath{M_{\mathrm{ W }}}}
\newcommand{\GW}{\ensuremath{\Gamma_{\mathrm{ W }}}}

\newcommand{\RunZ}{\hbox{Run-0}}
\newcommand{\RunIa}{\hbox{Run-Ia}}
\newcommand{\RunI}{\hbox{Run-I}}

\newcommand{\RunII}{\hbox{Run-II}}

\newcommand{\bluealle}{19}

\newcommand{\syspscalecombin}{7}

\newcommand{\sysholecombin}{2}

\newcommand{\sysbkgcombin}{3}
\newcommand{\sysgtwocombin}{5}
\newcommand{\sysqedcombin}{4}
\newcommand{\syspdfcombin}{10}
\newcommand{\statcombin}{12}

\begin{document}
\begin{center}
{\large \bf FERMI NATIONAL ACCELERATOR LABORATORY}
\end{center}

\baselineskip 20 pt

\begin{flushright}
       TEVEWWG/WZ 2012/01 \\
FERMILAB-TM-2532-E \\
       CDF Note 10812 \\
       D0 Note 6319 \\[1mm]
       30$^{\rm th}$ March 2012\\
\end{flushright}

\vskip 1cm

\begin{center}
{\Large\bf 2012 Update of the Combination of CDF and D0 Results 
                  for the Mass of the $W$ Boson}
\vfill

{\Large
The Tevatron Electroweak Working Group\footnote{
The Tevatron Electroweak Working group can be contacted at tev-ewwg@fnal.gov.\\
 \hspace*{0.20in} More information is available at {\tt http://tevewwg.fnal.gov}.}\\
for the CDF and D0 Collaborations}

\vfill

{\bf Abstract}

\end{center}

{ We summarize and combine the results on the direct measurements of
  the mass of the $W$ boson in data collected by the Tevatron
  experiments CDF and D0 at Fermilab. Earlier results from CDF {\RunZ}
  (1988--1989), D0 and CDF {\RunI} (1992--1995) and D0 results from
  1~fb$^{-1}$ (2002--2006) of {\RunII} data are now combined with two
  new, high statistics {\RunII} measurements: a CDF measurement in
  both electron and muon channels using 2.2 fb$^{-1}$ of integrated
  luminosity collected between 2002 and 2007, 
  and a D0 measurement in the electron channel using 4.3
  fb$^{-1}$ collected between 2006 and 2009.  As in previous combinations, the
  results are corrected for  inconsistencies in parton distribution
  functions and assumptions about electroweak parameters used in the
  different analyses.  The resulting Tevatron average for the mass of
  the $W$ boson is $\MW = 80,387\pm 16~\MeV$ and a new world average
  including data from LEP II is $\MW = 80,385\pm 15~\MeV$.}

\vfill



\section{Introduction}

The CDF and D0 experiments at the Tevatron proton-antiproton collider
located at the Fermi National Accelerator Laboratory have made several
direct measurements of the width, $\GW$, and mass, $\MW$, of the $W$
boson.  These measurements use both the $e\nu_e$ and $\mu\nu_\mu$
decay modes of the $W$ boson.

Measurements of $\MW$ have been published by CDF from the data of
{\RunZ}~\cite{MW-CDF-RunZ}, {\RunI}~\cite{MW-CDF-RUN1A, MW-CDF-RUN1B}
{\RunII}~\cite{MW-CDF-RUN2a} and by D0 from
{\RunI}~\cite{MW-D0-Ia,MW-D0-I,MW-D0-I-rap,MW-D0-I-edge} and
{\RunII}~\cite{MW-D0-RUN2a}. This document adds new {\RunII}
measurements from CDF~\cite{MW-CDF-RUN2} and
D0~\cite{MW-D0-Run2b}. The new CDF result supersedes and replaces the
 $200$ pb$^{-1}$ {\RunII} measurement~\cite{MW-CDF-RUN2a} used
in previous combinations.  There are no new measurements of the width
of the $W$ since the previous average~\cite{WID10} in March 2010.

This note reports the combination of the mass measurements. The
combination takes into account the statistical and systematic
uncertainties as well as correlations among systematic
uncertainties. It supersedes the previous
combinations~\cite{MWGW-RunI-PRD,TEV08, TEV09}. As with previous
combinations, the measurements are combined using the analytic BLUE
method~\cite{BLUE}.
\section{New measurements}

\subsection{CDF}

The CDF collaboration has extended its
measurement~\cite{MW-CDF-RUN2} of the $W$ boson mass to use $2.2$
fb$^{-1}$ of {\RunII} data taken between 2002 and 2007. Both the muon
($\mu\nu_\mu$) and electron ($e\nu_e$) channels are included and the
central tracker has been used to set the absolute energy scale, using
$J/\psi$ and $\Upsilon$ decays in addition to the $Z$ boson.  The CDF
result comes from the combination of 6 observables, $\mtmu$, $\ptmu$,
$\METmu$, $\mte$, $\pte$ and $\METe$.  The combined result is $\MW =
80,387 \pm 12~\text{(\rm stat.)} \pm 15~\text{(\rm syst.)}$ MeV.   
 Table
\ref{table:cdferr} summarizes the sources of uncertainty in the CDF
measurement.

\begin{table}[!hbp]
\begin{center}
\begin{tabular}{|l|c|}
\hline
Source & Uncertainty (MeV)   \\ [0.05in]
\hline \hline
Lepton energy scale and resolution             & \syspscalecombin\  \\
Hadronic recoil energy scale and resolution            &   6 \\
Lepton removal            &    \sysholecombin\    \\
Backgrounds               &    \sysbkgcombin\      \\
\hline
Experimental subtotal  & 10 \\
\hline
Parton distributions      &   \syspdfcombin\   \\
QED radiation             &  \sysqedcombin\    \\
$p_T(W)$ model  &  \sysgtwocombin\  \\ \hline 
Production subtotal &12\\ \hline
  Total systematic uncertainty                     & 15\\
\hline \hline
$W$-boson statistics     &    \statcombin\     \\
\hline \hline
Total  uncertainty                   &    \bluealle\     \\
\hline
\end{tabular}
\end{center}
\caption{Uncertainties for the combined result on $M_W$ from CDF~\cite{MW-CDF-RUN2}.}
\label{table:cdferr}

\end{table}

\subsection{D0}

The D0 collaboration has measured the $W$ boson
mass~\cite{MW-D0-Run2b} using $4.3$ fb$^{-1}$ of {\RunII} data taken
between 2006 and 2009.  The measurement is performed in the electron
channel and uses the $Z\rightarrow e^+e^-$ process as the 
calibration for the energy scale, effectively measuring the ratio
$M_W/M_Z$.  This calibration method eliminates many common systematic
uncertainties but is limited by the available $Z$ boson statistics.
This measurement is complementary to the previous
{\RunII} measurement~\cite{MW-D0-RUN2a} which used data from
2002--2006.

 \begin{table}[!hbp]
\begin{center}

\begin{tabular}{|l|c|}
\hline
Source & Uncertainty (MeV)   \\ [0.05in]

  \hline \hline
  Electron energy calibration       & 16 \\
  Electron resolution model        &  2 \\
  Electron shower modeling          &  4 \\
  Electron energy loss model     &  4 \\
  Hadronic recoil energy scale and resolution     & 5 \\
  Electron efficiencies      &  2 \\
  Backgrounds                  &  2 \\ \hline
  Experimental subtotal          & 18 \\ \hline
				    				     
  Parton distributions              & 11 \\
  QED  radiation               &  7 \\
  $p_T(W)$ model          &  2 \\ \hline
  Production subtotal       &  13\\ \hline
\hline
  Total systematic uncertainty                     & 22 \\
  \hline
\hline
  $W$-boson statistics & 13\\
  \hline \hline
  Total uncertainty &26 \\
  \hline
  \end{tabular}
\end{center} 
\caption{Uncertainties for the new D0 $M_W$  measurement~\cite{MW-D0-Run2b}. \label{t:syst}}
\label{table:d0err}
\end{table}    

The new D0 result uses the two most sensitive observables: the
transverse mass $\mte$ and the electron transverse momentum $\pte$.
The missing transverse momentum~$\METe$ has larger uncertainties and
was not used in the combination. These observables yield a combined
$W$ boson mass, $\MW = 80,367 \pm 13~\text{(stat.)} \pm
22~\text{(syst.)}$~MeV.  
The individual contributions to the uncertainty
are summarized in Table \ref{table:d0err}.

\section{Combination with older Tevatron measurements}

As documented in the 2008 and 2009 combinations~\cite{TEV08,TEV09},
the oldest measurements from \RunZ\ and \RunIa\ were made using older
PDF sets and have been corrected~\cite{TEV08} to modern PDF sets.  These older results have
also been adjusted to use the same combination technique (``BLUE'') as in later combinations.
All masses have also been corrected for a small dependence on the input
$W$ boson width in order to achieve consistency across all
results. The value of \GW\ quoted here corresponds to a Standard Model
definition based on a Breit-Wigner propagator in the ``running-width
scheme", $1/{(M^2-\MW^2+iM^2\GW/\MW)}$, with a width parameter, $\GW =
2092.2 \pm 1.5 $ MeV predicted by the Standard Model ~\cite{PR} using
a revised 2012 world average $W$ boson mass of $80,385\pm 15 $ MeV.
All measured masses are corrected to this value using $\Delta M_W =
-(0.15\pm0.05) \Delta\GW$ as in Ref.~\cite{TEV08}.  The $W$ boson mass
uncertainty arising from an uncertainty in the $W$ boson width is now
consistently treated across all measurements.

Table \ref{tab:MW-inputs12} summarizes all of the inputs to the
combination and the corrections made to ensure consistency across
measurements.

\begin{table}[!hbp]
\begin{center}
\begin{tabular}{|l |r |r |r |r |r |r |r|r|} \hline

  & \text{CDF-0} & \text{CDF-Ia} & \text{CDF-Ib} & \text{D0-I} &
  \text{CDF-II} & \text{D0-II} & \text{D0-II} \\
  &  &  &  & & {\small (2.2 fb$^{-1}$)} & \small{(1.0 fb$^{-1}$)} & \small{(4.3 fb$^{-1}$)} \\
 \hline
 \text{$M_W$ published} & 79910 & 80410 & 80470 & 80483 & 80387 & 80400.7 & 80367.4 \\
 \text{Uncertainty (publ.)} & 390 & 180 & 89 & 84 & 19 & 43 & 26 \\
 \text{$\Gamma_W$ (publ.)} & 2100 & 2064 & 2096 & 2062 & 2094 & 2099.6 & 2100.4 \\
 \hline
 {\bf Corrections}&&&&&&&\\
 \text{$\Delta \Gamma $} & 1.2 & $-$4.2 & 0.6 & $-$4.5 & 0.3 & 1.1 & 1.2 \\
 \text{PDF} & 20 & $-$25 & 0 & 0 & 0 & 0 & 0 \\
 \text{BLUE} & $-$3.5 & $-$3.5 & $-$0.1 & 0 & 0 & 0 & 0 \\
 \text{Total} & 17.7 & $-$32.7 & 0.5 & $-$4.5 & 0.3 & 1.1 & 1.2 \\
 \hline
 \text{$M_W$(corr)} & 79927.7 & 80377.3 & 80470.5 & 80478.5 & 80387.3 & 80401.8 & 80368.6 \\
 \hline
 {\bf Uncertainties}&&&&&&&\\
 \text{Total} & 390.9 & 181.0 & 89.3 & 83.4 & 19.0 & 43.2 & 25.8 \\
 \text{PDF} & 60 & 50 & 15 & 8.1 & 10& 10.4 & 11 \\
 \text{Rad. Corr.} & 10 & 20 & 5 & 12 & 4 & 7.5 & 7 \\
 \text{$\Gamma_W$} & 0.5 & 1.4 & 0.3 & 1.5 & 0.2 & 0.4 & 0.5\\
 \hline \hline
\end{tabular}
\caption{The inputs used in the \MW\ combination. All entries are in MeV.
   \label{tab:MW-inputs12}}
\end{center}
\end{table}

\section{Correlation of the new Run II results with other Tevatron
  measurements}

The greatly improved statistical power of these recent measurements
has made systematic uncertainties in the $W$ boson production and
decay model  more significant.  As a result, a more careful
treatment of model uncertainties has been undertaken.  We have adopted a
standard in which, for shared model uncertainties, the minimum
uncertainty across experiments is assumed to be a fully correlated
uncertainty while excesses above that level are generally
assumed to be due to uncorrelated differences between experiments.
One exception is the two D0 Run-II measurements  which use very similar
models and are treated as fully correlated.

The experimental systematic uncertainties for D0 are dominated by the
energy scale for electrons and are almost purely
statistical, as they are  derived from the limited sample of
$Z$ decays.  CDF uses independent data from the central tracker to
set the muon and electron energy scales. For these reasons all of the experimental
uncertainties are assumed to be completely uncorrelated with each
other and with all previous measurements.

Three sources of systematic uncertainties due to modeling of the
production and decay of $W$ and $Z$ bosons are assumed to be at least
partially correlated between all Tevatron measurements, namely (1) the
parton distribution functions (PDFs), (2) the assumed width of the $W$
boson ($\Gamma_W$) and (3) the electroweak radiative
corrections. 
 
 \begin{enumerate}
 \item {\underline{PDFs}}\\
   Both experiments use the CTEQ6.6~\cite{cteq66} parton distribution
   as the central PDF set in their $W$ boson production model.  D0
   uses the older CTEQ6.1~\cite{CTEQ61M} error set to estimate the PDF
   uncertainties while CDF uses MSTW2008~\cite{mstw2008} and
   cross-checks with the CTEQ6.6 error set. Because these PDF sets are
   similar, and rely on common inputs, the uncertainties introduced by
   PDFs in the recent measurements are assumed to be largely
   correlated and treated using the prescription for partial
   correlations described above. \footnote{The D0 \RunI\ measurement
     used both central and forward electrons. Because of the larger
     rapidity coverage, the PDF uncertainty was both smaller and
     different from all other measurements considered here.
     Simulation studies indicate a correlation of approximately 70\%
     between the PDF component of this measurement and the others.}
 \item {\underline{Width of the $W$ boson}}\\
   We assume that the very small uncertainty due to the uncertainty in
   the $W$ boson width
   is 100\% correlated across all measurements.
 \item {\underline{Radiative Corrections}}\\
   Current estimates of the uncertainties due to radiative corrections
   include a significant statistical component.  The {\sc
     PHOTOS}~\cite{ref:PHOTOS} radiative correction model has been
   used  in the recent measurements with cross-checks from
   {\sc W(Z)GRAD}~\cite{ref:WGRAD} and {\sc
     HORACE}~\cite{horace}. These studies yield model differences
   consistent within statistical uncertainties. We assume that there is a correlated
   uncertainty of 3.46 MeV\cite{MW-CDF-RUN2} due to the common use of PHOTOS with the
   remaining uncertainties being uncorrelated.
 
 \end{enumerate}
 
\section{Combination of Tevatron $\MW$ measurements}
\label{cov}

The seven measurements of $\MW$ to be combined are given in
Table~\ref{tab:MW-inputs12} and include both new results discussed
above but exclude the superseded 0.2~fb$^{-1}$ CDF \RunII\
result~\cite{MW-CDF-RUN2a}.  The combined Tevatron $W$ boson mass,
from all measurements, calculated using the BLUE method~\cite{BLUE}
is:
\begin{equation}
M_{W} = 80,387 \pm 16 \ \textrm{MeV}\,.
\end{equation}
The $\chi^{2}$ for the combination is of 4.2 for 6 degrees of freedom, with a
probability of 64\%.  Table \ref{contribution} shows the relative weights of
each measurement in the combination. 
\begin{table}
\begin{center}
\begin{tabular}{|c |c |}  \hline 
           &    Relative Weights in \%  \\ \hline \hline
CDF-0      &   0.1  \\
CDF-Ia     &   0.5  \\
CDF-Ib     &   1.9  \\
D0-I     &  2.8  \\
CDF-II {(2.2 fb$^{-1}$) }   &  60.3  \\ 
D0-II {(1.0 fb$^{-1}$)}   &  7.9 \\
D0-II {(4.3 fb$^{-1}$)}   &  26.5  \\ \hline \hline
\end{tabular}
\caption{Relative weights of the contributions in \%. \label{contribution}}
\end{center}
\end{table}

The
global correlation matrix for the seven measurements is shown in
Table~\ref{global}.

\section{World Average}

\begin{table}
\begin{center}
\small{
\begin{tabular}{|l |l |l |l |l |l |l |l |l  } \hline
  & CDF-0  &CDF-Ia    &CDF-Ib   & D0-I   &CDF-II 
   &D0-II &D0-II \\
& & & && {(2.2 fb$^{-1}$)} & {(1.0 fb$^{-1}$)} & {(4.3
      fb$^{-1}$)} \\
\hline  \hline
CDF-0                    & 1. & 0.002 & 0.003 & 0.002 & 0.015 & 0.007 & 0.011 \\
CDF-Ia                      &0.002 & 1. & 0.007 & 0.005 & 0.033 & 0.014 & 0.024 \\
CDF-Ib                  &0.003 & 0.007 & 1. & 0.009 & 0.066 & 0.029 & 0.049 \\
D0-I                &0.002 & 0.005 & 0.009 & 1. & 0.044 & 0.019 & 0.032 \\
CDF-II {(2.2 fb$^{-1}$) } &0.015 & 0.033 & 0.066 & 0.044 & 1. & 0.137 & 0.23 \\
D0-II {(1.0 fb$^{-1}$)}   &0.007 & 0.014 & 0.029 & 0.019 & 0.137 & 1. & 0.137 \\

D0-II {(4.3 fb$^{-1}$)}   &0.011 & 0.024 & 0.049 & 0.032 & 0.23 & 0.137 & 1.\\ \hline
\end{tabular}
}
\caption{Correlation coefficients between the different experiments. \label{global}}
\end{center}
\end{table}
We also combine all of the Tevatron measurements with the $W$ boson mass determined from $WW$ production at LEP II~\cite{LEPEWWG07}.
The combination of all of the Tevatron results with the LEP II preliminary
result of $80,376 \pm 33$ MeV~\cite{LEPEWWG07}, assuming no correlations, yields a new world average for the $W$ boson mass:
\begin{equation}  
M_{W} = 80,385 \pm 15 \ \textrm{MeV}\,.  
\end{equation}

\begin{figure}[hbt]
\centering \includegraphics [width=.90\textwidth] {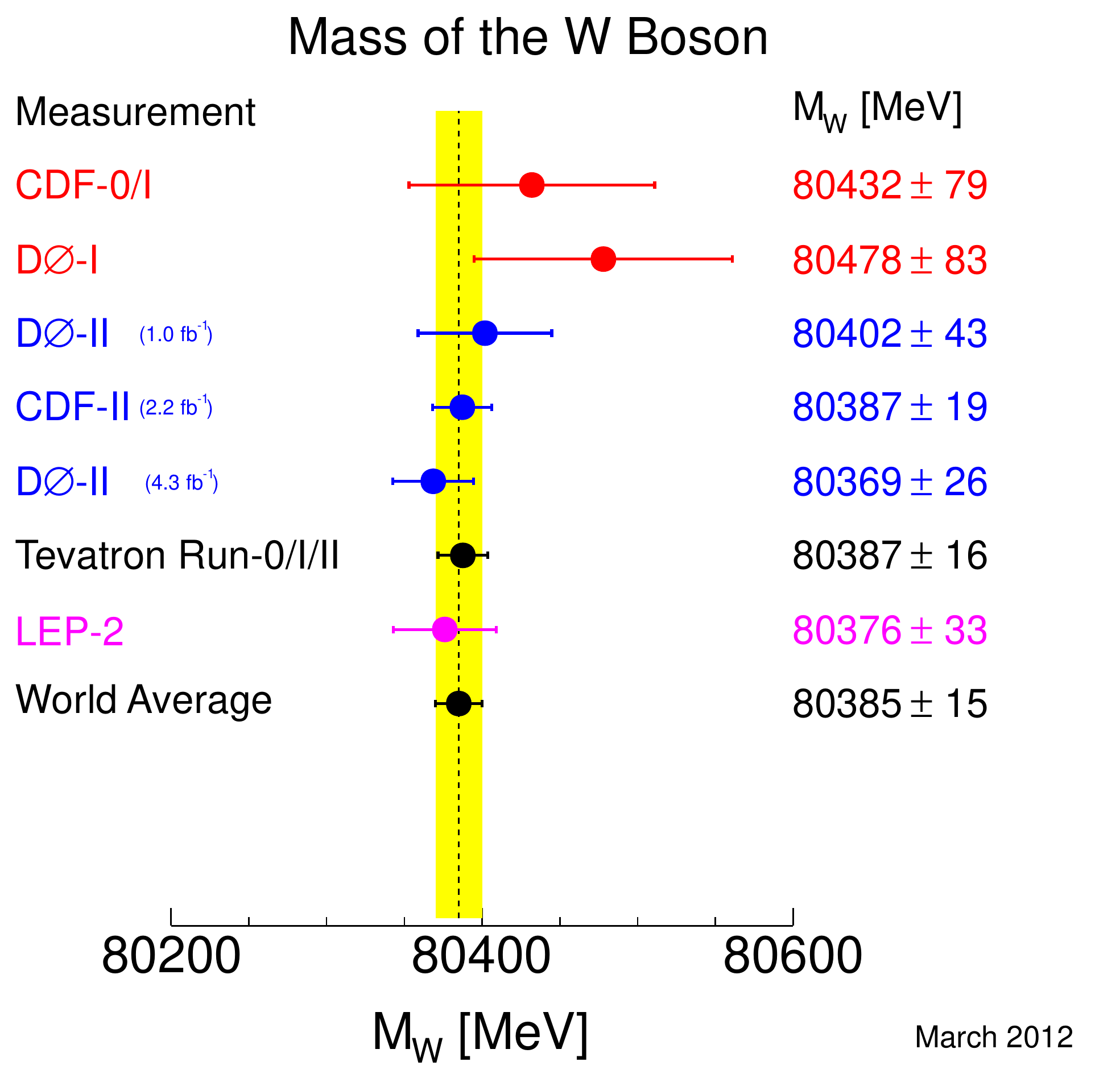}
\caption{ Summary of the measurements of the $W$ boson mass and their
  average as of March 2012. The result from the Tevatron corresponds
  to the values in this note (see Table \ref{tab:MW-inputs12}) which
  include corrections to the same $W$ boson width (2092.2 MeV) and PDFs.  The LEP
  II result is from Ref.~\cite{LEPEWWG07}. An estimate of the world
  average of the Tevatron and LEP results assuming no correlations
  between the Tevatron and LEP is included.}
\label{fig:sum}
\end{figure}

The $\chi^2$ is 4.3 for 7 degrees of freedom with a probability of 74\%.
Figure~\ref{fig:sum}  illustrates the new combined $W$ boson mass estimates.

\clearpage
 

\end{document}